\documentclass[lettersize,journal]{IEEEtran}
\usepackage{amsmath,amsfonts}
\usepackage{algorithmic}
\usepackage{algorithm}
\usepackage{array}
\usepackage[caption=false,font=normalsize,labelfont=sf,textfont=sf]{subfig}
\usepackage{textcomp}
\usepackage{stfloats}
\usepackage{url}
\usepackage{verbatim}
\usepackage{graphicx}
\usepackage{cite}
\usepackage{multirow}
\usepackage{booktabs}
\begin{document}


\title{Experimental Demonstration of Partially Disaggregated Optical Network Control\\Using the Physical Layer Digital Twin}

\author{\vspace{6mm}Giacomo Borraccini,~\IEEEmembership{Graduate Student Member,~IEEE}, Stefano Straullu, Alessio Giorgetti,\\Renato Ambrosone, Emanuele Virgillito,~\IEEEmembership{Graduate Student Member,~IEEE},\\Andrea D'Amico,~\IEEEmembership{Graduate Student Member,~IEEE}, Rocco D'Ingillo,~\IEEEmembership{Graduate Student Member,~IEEE}, Francesco Aquilino, Antonino Nespola, Nicola Sambo, Filippo Cugini, and Vittorio Curri,~\IEEEmembership{Senior Member,~IEEE}
\thanks{Giacomo Borraccini, Renato Ambrosone, Emanuele Virgillito, Andrea D'Amico, Rocco D'Ingillio and Vittorio Curri are with the Department of Electronics and Telecommunications~(DET), Politecnico di Torino, Turin 10129, Italy (email: giacomo.borraccini@polito.it, renato.ambrosone@polito.it, emanuele.virgillito@polito.it, andrea.damico@polito.it, rocco.dingillo@polito.it, vittorio.curri@polito.it).}%
\thanks{Renato Ambrosone is also with GARR Consortium, Rome 00185, Italy.}
\thanks{Stefano Straullu, Francesco Aquilino and Antonino Nespola are with LINKS Foundation, Turin 10138, Italy (email: stefano.straullu@linksfoundation.com, francesco.aquilino@linksfoundation.com, antonino.nespola@linksfoundation.com).}%
\thanks{Alessio Giorgetti is with IEIIT-CNR, Pisa 56122, Italy (email: alessio.giorgetti@ieiit.cnr.it).}%
\thanks{Nicola Sambo is with the Scuola Superiore Sant'Anna, Pisa 56127, Italy (email: n.sambo@sssup.it).}%
\thanks{Filippo Cugini is with the CNIT, Pisa 56124, Italy (email: filippo.cugini@cnit.it).}}

\markboth{Journal of \LaTeX\ Class Files,~Vol.~14, No.~8, August~2021}%
{Shell \MakeLowercase{\textit{et al.}}: A Sample Article Using IEEEtran.cls for IEEE Journals}


\maketitle

\begin{abstract}
Optical communications and networking are fast becoming the solution to support ever-increasing data traffic across all segments of the network, expanding from core/metro networks to 5G/6G front-hauling.
Therefore, optical networks need to evolve towards an efficient exploitation of the infrastructure by overcoming the closed and aggregated paradigm, to enable apparatus sharing together with the slicing and separation of the optical data plane from the optical control.
In addition to the advantages in terms of efficiency and cost reduction, this evolution will increase the network reliability, also allowing for a fine trade-off between robustness and maximum capacity exploitation.
In this work, an optical network architecture is presented based on the physical layer digital twin of the optical transport used within a multi-layer hierarchical control operated by an intent-based network operating system.
An experimental proof of concept is performed on a three node network including up to 1000~km optical transmission, open re-configurable optical add \& drop multiplexers (ROADMs) and white-box transponders hosting pluggable multi-rate transceivers.
The proposed solution is based on GNPy as optical physical layer digital twin and ONOS as intent-based network operating system.
The reliability of the optical control decoupled by the data plane functioning is experimentally demonstrated exploiting GNPy as open lightpath computation engine and software optical amplifier models derived from the component characterization.
Besides the lightpath deployment exploiting the modulation format evaluation given a generic traffic request, the architecture reliability is tested mimicking the use case of an automatic failure recovery from a fiber cut.
\end{abstract}


\section{Introduction}

\IEEEPARstart{S}{ervice} providers and network operators are showing interest in disaggregated optical networks, vendor-neutral control and management and multi-vendor interoperability as a way to overcome vendor lock-in and save capital outlay~\cite{facebook, riccardi-jlt18}.
One specific example is the separation of control, data, and management planes, which has become increasingly important in recent years.
Disaggregation may help operators to overcome vendor-lock-in at the control plane level when used in conjunction with the specification of control and management of transponders, transceivers (TRXs), and re-configurable optical add \& drop multiplexers (ROADMs).
A considerable amount of effort is going into creating standardized data models that vendors and operators can all utilize.
The Yet Another Next Generation (YANG)~\cite{yang,dallaglio-jocn17} data modeling has emerged as the preferred language for interacting with the control and management system.
The Network Configuration (NETCONF) protocol, which was established by the Internet Engineering Task Force, supports YANG (IETF)~\cite{rfc-netconf}.
Open and disaggregated networking is being defined by a number of consortia and initiatives, including OpenConfig~\cite{openconfig}, OpenROADM~\cite{open-roadm}, and Telecom Infra Project (TIP)~\cite{tip}, which include major operators, service providers, and manufacturers.
For the definition of transponders, the OpenConfig consortium has given the YANG data format~(e.g.,~\cite{t600}) and various solutions on the market currently support the OpenConfig YANG data model.
In addition, there are several works in the literature that show how to configure network elements (NEs) or subsystems using NETCONF/YANG, transport performance parameters and reconfigure these elements according to them.
With the aid of specially created data models, it has been demonstrated that NETCONF and YANG can configure and monitor transponders \cite{dallaglio-jocn17, sambo-jlt19, sambo-trial19}.
The authors of~\cite{sgambelluri-access20} created two different types of agents: an OpenConfig-based transponder agent and an OpenROADM-based line system agent.
YANG is used in~\cite{velasco-jlt18} to enable device configuration, monitoring of operational data and physical parameters, local control loops for basic parameter reconfiguration/tuning and notifications incorporating alarms.
In~\cite{kundrat-jlt18}, the authors demonstrate disaggregation allowing automated optical path protection through a software-defined networking (SDN) controller.
In~\cite{nadal-jocn21}, a sliceable transponder supporting spectrum and space dimensions is supplied with an implementation of an OpenConfig-based agent (i.e., local controller of a device).
Telemetry applications based on the gRPC protocol~\cite{paolucci-jlt18} have also been modeled using YANG.
Exploiting NETCONF and other industry-standard protocols, the open ONOS controller in~\cite{campanella-ofc19} performs discovery and control of the network topology.
The quality of transmission (QoT) estimation is another feature that has to be addressed, particularly for lightpath (LP) provisioning and maintenance.
An LP is a transmission channel in the optical domain, defined by the physical path between a source and a destination node, which does not include any optical-electronic-optical conversion, and the frequency slot used, thus implying the wavelength continuity.
As far as coherent transmission technology is concerned, the ability to recover the constellation phase noise by means of carrier phase estimation (CPE) algorithms and the linear mapping between optical field and electric received signals allows to model the optical transmission as a an additive white and Gaussian noise (AWGN) channels, enabling optical transmission through transparent LPs using coherent technology.
In this condition, the generalized signal to noise ratio (GSNR)~\cite{filer2018multi} can be employed as a unique figure of merit for QoT.
The GSNR is determined by dividing the power of the channel being tested by the total of the accumulated amplified spontaneous emission (ASE) noise caused by optical amplifiers and the non-linear interference (NLI) impairment caused by fiber propagation.
The AWGN abstraction of LPs has been implemented in the TIP's GNPy, an open-source and vendor-neutral QoT estimator~\cite{tip, oopt-gnpy, curri-ofc21, curri2022gnpy}.
In order to enforce fundamental transmission parameters on transponders and ROADMs for LP provisioning or restoration, SDN controllers have so far used QoT estimators, even if further improvements are required to adequately handle the optical data plane technology's complexity.
%
%
For example, there are no erbium doped fiber amplifier (EDFA) drivers available in the ONOS repository, and existing ONOS implementations do not take into account these NEs, which impact significantly on the overall QoT.
Considering the market trend towards the deployment of optical multi-band transmission, expanding SDN controllers to include such optical amplifiers is a required feature to cope with scalability and performance challenges.
Moreover, the inclusion of open packet-optical nodes housing pluggable TRXs has an additional impact on complexity and resilience since it needs the SDN controller to operate on the setup of new parameters, such as at the packet level.

The authors experimentally verified the functioning of a partially disaggregated optical network with a linear topology (up to 1400~km of distance)~\cite{borraccini2022qot}.
Furthermore, the functioning of the physical layer digital twin and the optical network controller within the overall infrastructure orchestration has also been highlighted in~\cite{borraccini2023disaggregated} and the presented architecture has been implemented on an experimental multi-vendor triangular-topology system.

In this work, a partially disaggregated optical network architecture is systematically described and fully validated on the same multi-vendor triangular-topology network, controlling both open packet-optical nodes, i.e. packet switches hosting pluggable coherent TRXs, and a completely effective optical data plane encompassing ROADM whiteboxes and optical amplifiers.
The ONOS SDN controller exploits NETCONF protocol~\cite{onos, giorgetti2020control} with OpenConfig vendor-neutral YANG data models for the control of pluggables and ROADMs from multiple vendors.
In addition, ONOS relies on the GNPy for GSNR estimation~\cite{borraccini2021cognitive}.
The GNPy tool is also leveraged for the configuration of the optical amplifiers, guaranteeing full visibility and control of the analog parameters (e.g., gain profile) while freeing the SDN controller from the burden of controlling all network EDFAs.
In this way, the SDN controller can maintain visibility of network parameters, while the GNPy tool can practically evolve as a physical layer digital twin solution for the entire optical infrastructure.

The article consists of four main sections.
In Sect.~II, an architecture for the control of an open and disaggregated optical network is presented in terms of involved actors and their interactions, providing further details regarding the functioning of the optical control plane exploiting a cognitive approach, and of the data plane, in the respective subsections.
In Sect.~III, the experimental apparatus set up to validate the proposed network architecture is reported, focusing on the instrumentation, the various implementation choices and the adopted precautions.
In Sect.~IV, the experimental results obtained during the various steps of the validation are reported and commented on.
In particular, it is observed: the cognitive process of the control plane as regards the characterization of the physical layer and the optimization of the working point of the amplifiers, the transmission performance of the two paths present in the topology, the functioning of the deployment and recovery procedure of LPs with relative time lapse measurements.
Finally, in Sect.~V the conclusions are outlined.

\begin{figure*}[t]
\centering
\includegraphics[width=\linewidth]{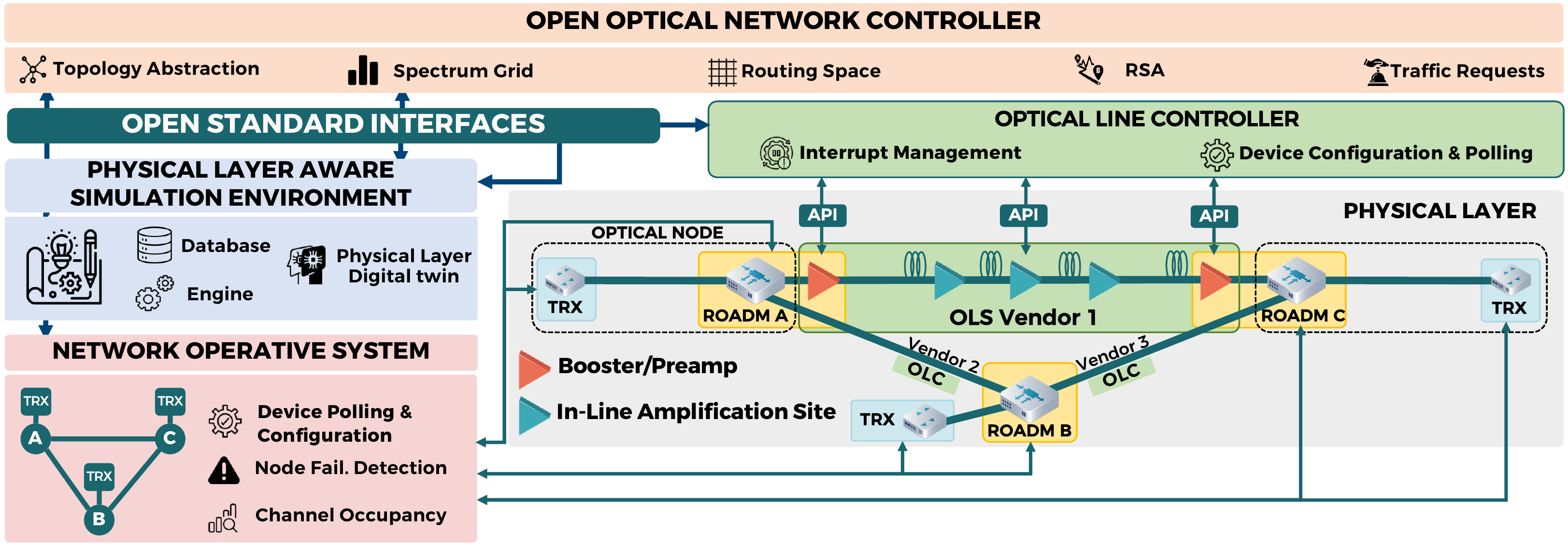}
\caption{Abstract scheme of the open and disaggregated optical network architecture, designed to independently manage the optical data and control plane.}
\label{fig:network_arch}
\end{figure*}

\section{Network Architecture}

The proposed optical network architecture is designed to work in an open and disaggregated context, aiming to achieve interoperability among multi-vendor equipment and decouple the optical data and control planes.
Specifically, the target approach is a partially disaggregated management~\cite{pham2020demonstration} in which each node-to-node optical link is typically provided by a single vendor.
Thus, the NEs of a line are meant to be handled by a single controller, collapsing the management to a single element from a network point of view.
The representation in Fig.~\ref{fig:network_arch} schematizes an abstract network structure.

Starting from the physical layer~(PHY), the optical hardware infrastructure is composed by TRXs, ROADM whiteboxes, optical fiber spools and optical amplifiers.
A couple of booster~(BST) and pre-amplifier~(PRE) is assumed to be integrated in a single ROADM for each switching direction.
In this network architecture, a single optical line system~(OLS) is defined as a ROADM-to-ROADM optical line, thus including the BST and the PRE of the ROADMs at both line terminals.
Each amplification site along an OLS can host amplifiers of different technologies, e.g. erbium-doped fiber amplifiers~(EDFAs) or Raman amplifiers, and telemetry devices, such as an optical time domain reflectometer~(OTDR), photo-diodes and optical channel monitors~(OCMs).
An optical node consists of the set of TRXs and ROADMs placed in a specific location within the geographical footprint of the optical network.

The optical equipment is managed by the cooperation of four different software modules: (i) the network operative system~(NOS); (ii) an optical line controller~(OLC) for each OLS composing the optical network; (iii) the physical layer aware simulation environment~(PLASE); and (iv) the open optical network controller (OONC).
The communication among the modules, TRXs and ROADMs is performed exploiting open standard interfaces and protocols (REST, NETCONF, etc..).

The NOS is aware of the status of the optical nodes and their connections, building an abstraction of the network topology and directly controls each TRX and ROADM within the network.
Furthermore, the NOS must properly manage different types of failure detected at the PHY, both from nodes and lines.
Each OLC is responsible for the management of the corresponding OLS, which is generally provided by a specific vendor.
The OLC communicates through defined application programming interfaces~(APIs) with all the amplification sites of the OLS and the BST and the PRE, collecting telemetry information by means of device polling, configuring the amplifiers' working point and notifying status information derived from interrupt management.
%
%
The PLASE represents the building block of the network architecture within which all the intelligence regarding the PHY is collected.
In particular, the PLASE stores data related to the PHY, such as datasets or amplifier/ROADM/TRX/fiber characterizations.
The digital twin of the physical layer (PHY-DT) represents the central element of the PLASE, including the model of the physical layer and allowing to simulate the behavior of the system.
On top of this, several computational algorithms are run to fulfill different tasks which rely on the PHY information knowledge.
The PLASE directly communicates both with the OLCs and the OONC.
With respect to the OLCs, the PLASE collects telemetry data from the NEs, performing fundamental operations such as the working point optimization of a specific OLS and the LP computation engine~(L-PCE)~\cite{9382009}.

The OONC implements the system north-bound interface exposed towards the network users.
This module orchestrates the deployment process, for instance, it transparently realize the LP allocation and recovery in the optical network, interacting with the other software modules (i.e., the NOS and the PLASE).
Specifically, the OONC constructs the description of the spectrum grid by supplying a single configuration for the active channel frequencies via a static external configuration recording the center frequencies for all optical DWDM signals.
As a result of the topology abstraction offered by the NOS, the routing space is then created and contains details on all conceivable pathways inside the network as well as the availability of the channel wavelength along each path.
By making use of such structures, the OONC carries out the routing spectrum assignment (RSA), helped by the PLASE transmission performance indicators, in response to incoming traffic requests from outside pairs of source-destination nodes.
In the following sections, the behaviour of the proposed network architecture is illustrated, focusing individually on the optical control and data plane operations and showing how, in our perspective, such two aspects may be decoupled and independently managed from a procedural point of view.

\begin{figure*}[t]
\centering
\includegraphics[width=0.88\linewidth]{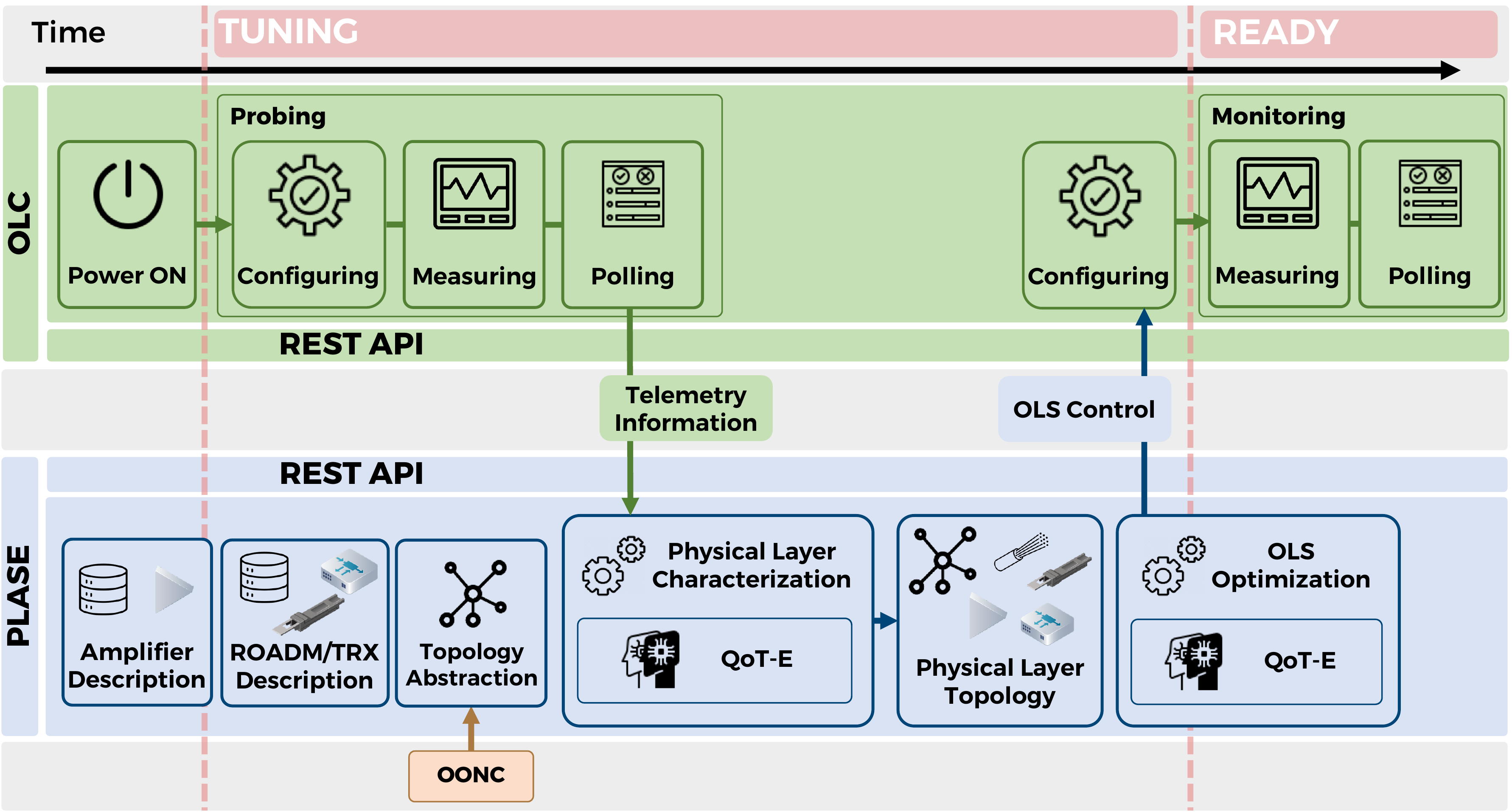}
\caption{Qualitative time line of the proposed cognitive optical control plane operation that leads an OLS to be ready for use.}
\label{fig:cognitive}
\end{figure*}

\subsection{Cognitive Optical Control Plane}

The optical control plane is in charge to manage the optical equipment in order to maximize the exploitation of the installed resources in terms of transmission performance.
This task requires the possibility to choose the most advantageous strategy for the amplifiers' working point setting, while minimizing the allocated margins~\cite{pointurier2017design}.
The latter condition is directly related to the degree of knowledge of the PHY's devices, both lumped (optical switches, connectors, EDFAs) and components inducing distributed effects such as fibers or Raman amplifiers.
In particular, the effect of lumped losses and fibers can be appreciated only in field by means of the indirect estimation of their properties.
In this perspective, the network provisioning can be performed downstream of a \emph{probing} procedure using the available monitors and telemetry in order to completely characterize the PHY, in order to reduce the margin allocation.
The \emph{cognitive} property is the main feature of a class of optical networks bearing the same name~\cite{zervas2010cognitive, wei2012cognitive}.
In the network context described above, the optical control plane is represented by the collection of all the OLCs supervised by the PLASE.
In the following, the procedural steps that bring to the network provisioning are described focusing on the case of a single OLS~(Fig.~\ref{fig:cognitive}).
The procedure can be repeated over all the OLS, with no loss of generality, given the partial disaggregation context.
However, the process that leads the OLS to be ready to perform network operations is entirely dependent on the choices dictated by the OLS vendor, especially in the type of data that the OLC exposes.

After the installation and before starting the transmission operations, an OLS undergoes a tuning procedure targeting the definition of the working point of each amplifier along the line.
This procedure is meant as an automated probing of the OLS, aiming to achieve a more accurate knowledge of the PHY, capturing the behaviour of the system directly from the field and consequently adjusting the model adopted by the PHY-DT within the PLASE.
Such procedure is composed of three steps: 
\begin{itemize}
    \item Configuration of the equipment using predefined settings;
    \item Telemetry measurements of the quantities of interest;
    \item Device polling in order to collect the information.
\end{itemize}
A dedicated representational state transfer~(REST) API is used to transfer data from the OLC to the PLASE and vice versa.
The PLASE estimates the lumped losses and the properties of the fibers matching the collected measurements with the configuration of the equipment during the probing procedure, completing the characterization of the PHY.
At the end of the probing procedure led by an OLC, the related PHY descriptions of amplifiers, TRXs and ROADMs are stored in a static database shared with the PLASE.
\begin{figure*}[t]
\centering
\includegraphics[width=0.85\linewidth]{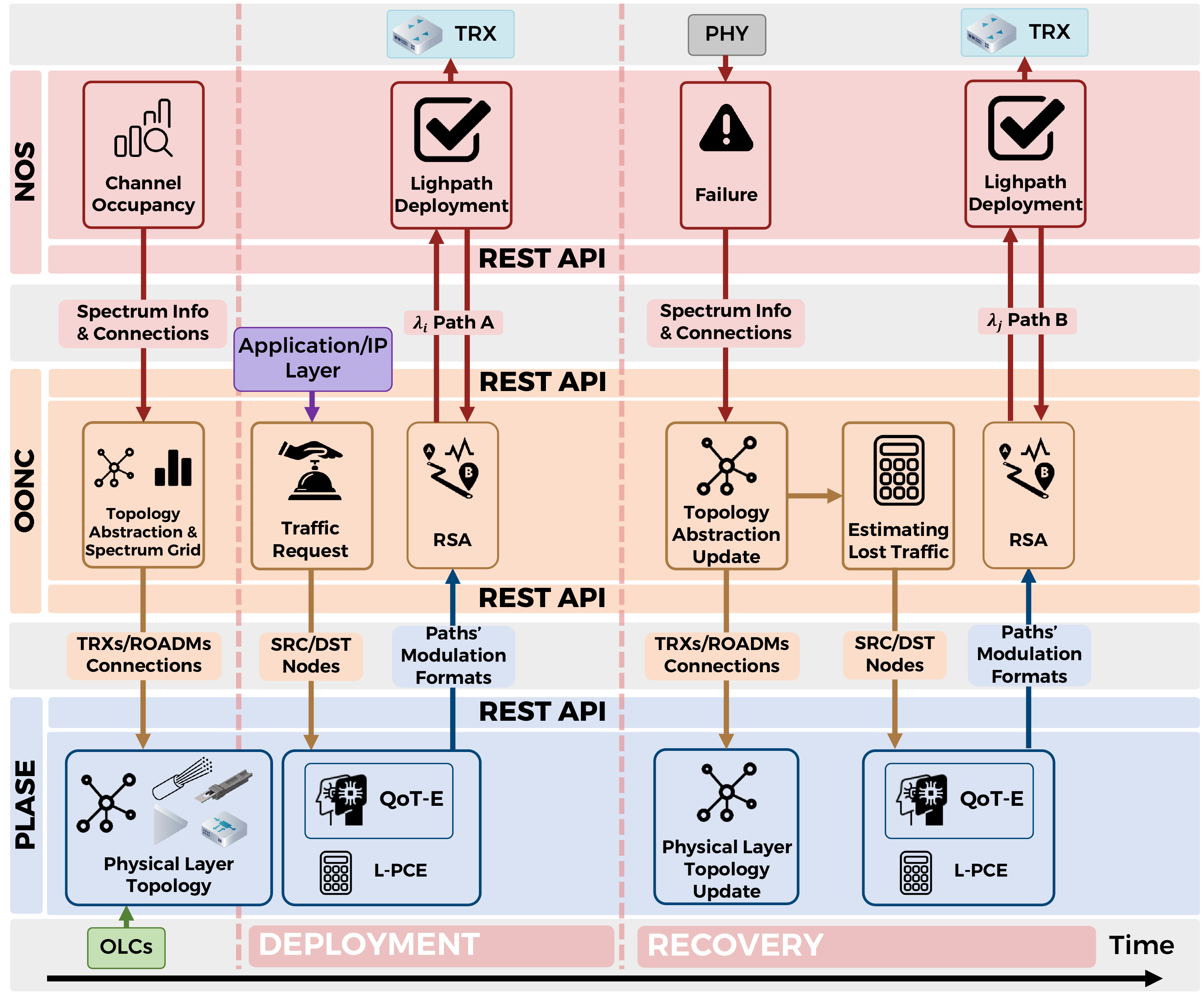}
\caption{Qualitative time line of the proposed optical data plane operation: establishment of a connection following a LP request and a recovery after a failure detection.}
\label{fig:data_plane}
\end{figure*}
On top of such a detailed PHY model, the OONC provides the topology abstraction, describing the available physical connections between TRXs and ROADMs, thus allowing the PLASE to identify the available OLS within the network.
Hence, the PLASE is able to build the complete network PHY topology retrieving for each OLS the corresponding virtualization combining the virtual topology, the data measured by the telemetry during the probing procedure and the PHY descriptions.
The PLASE defines the settings of each amplifier within the specific OLS on the bases of the physical layer topology in order to optimally match the transmission strategy.
The devised architecture assumes full spectral load operation at the basis of the optical control.
This corresponds to define the working point with respect to the worst case scenario in terms of transmission performance.
This assumption allows to decouple the operation of the data and control planes at an operational level, ensuring that the evaluation of the maximum modulation format cardinality for a specific path and wavelength is conservative even after the deployment of further LPs.
Once the configuration is received, the OLS is ready to operate as soon as the OLC sets the amplifiers at the design working point.
During the operations, the OLC periodically monitors the OLS status through the polling of the telemetry devices.

\subsection{Optical Data Plane Operation}

The optical data plane is in charge of managing the allocation of optical connections, or LPs, between couples of source-destination (SRC-DST) nodes based on the traffic requests coming from the application/IP layer.
The choice on the LP's allocation is performed defining the characteristics of the optical tributary signals in terms of occupied wavelength and modulation format, and then arranging the corresponding assigned path along the optical network~\cite{curri2022gnpy}.
Furthermore, a robust management of the previously allocated LPs has to be guaranteed, both fulfilling new requests and intercepting or possibly predicting failures, in order to promptly take countermeasures against a possible LP out-of-service.
In the described context, the operations of the optical data plane are entrusted to the OONC supported by the PHY-DT which performs the QoT estimation.
In this work, the focus is, first, on the deployment of a set of LPs between a couple of nodes to satisfy a given generic traffic request.
Secondly, the management of the traffic recovery is addressed, providing for the automatic re-establishment of the lost connections in case of hard failures.

The proposed qualitative time line of the operations is depicted in Fig.~\ref{fig:data_plane}.
During the network provisioning, the NOS establishes a communication with all the network optical nodes in order to identify which are the available physical connections between the nodes and retrieve the status of the current spectral occupancy in terms of already deployed LPs.
The OONC receives this information from the NOS and consequently draws a network topology abstraction and the initialization of the routing space on the basis of the acquired spectrum grid.
The collection of such physical TRX/ROADM connections is sent to the PLASE which are integrated with its OLS knowledge to create the complete picture of the PHY topology.
Once all the OLSs are ready to use, the deployment of a connection is engaged by a generic traffic request to the OONC, specifying the SRC/DST node couple and the required bit-rate.
Then, the nodes' pair is forwarded to the PLASE which calculates the available maximum modulation format through the PHY-DT for all the channels (wavelengths) and for all the physical paths having the request nodes as end points, by executing the L-PCE on the basis of the PHY topology and the TRX characterizations.
The OONC receives the evaluated modulation formats of all the available paths and performs the RSA defining the characteristics and the needed number of the LPs to deploy in order to satisfy the traffic request.
In particular, as preliminary step, the current routing space is used to check the path feasibility for all the retrieved modulation formats, selecting only the available channels that ensure wavelength continuity on a certain path.
The complete definition of each LP during the application of the RSA can require constraints and specifications which depend on the adopted algorithm.
Finally, the LP is practically put into operation by properly configuring the TRXs operational mode and the ROADMs directly through the NOS with the proper interfaces.
At the end of the operation, the NOS reports the outcome (success/failure) to the OONC in order to correctly manage the traffic request.

Moving ahead into the time line shown in Fig.~\ref{fig:data_plane}, the process to counteract hard-failures is described in the following.
During the life-cycle of an optical network, indeed, several types of failures may occur (e.g., due to fiber cuts), causing LP out-of-service.
Focusing on hard failures due to fiber cut events, the detection of such events can be performed both at the OLS level -- with an interrupt received by the NOS through the OLC -- or at the node's level -- observing the TRX or ROADM telemetry stream directly at the NOS level.
In the proposed architecture, as first step, the NOS intercepts the failure coming from the PHY at the optical node side and the OONC proceeds modifying the network topology abstraction.
\begin{figure*}[t]
\centering
\includegraphics[width=\linewidth]{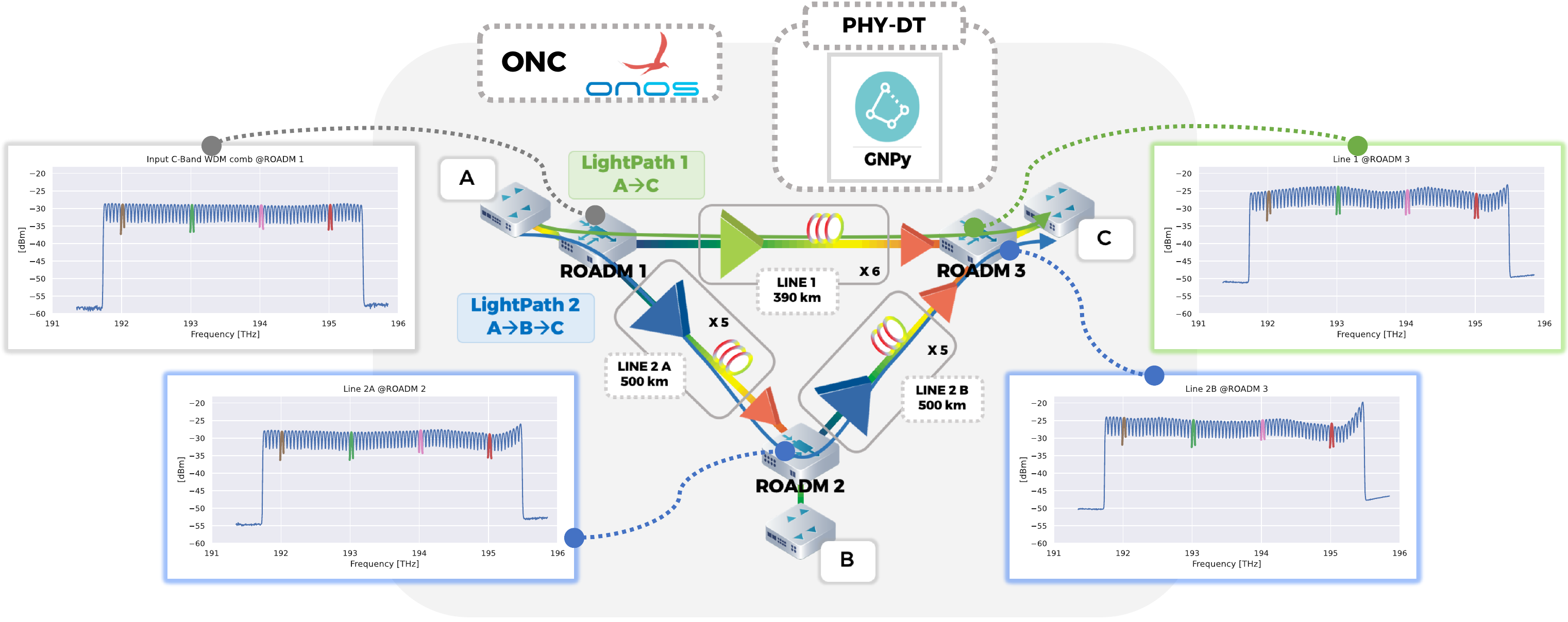}
\caption{Experimental setup of the optical network set in laboratory, including the spectra of the C-band WDM comb propagated through the various optical nodes. ONOS holds the role of NOS and GNPy represents the PHY-DT.}
\label{fig:setup}
\end{figure*}
This means that the failing link is removed from the available physical connection list of the topology abstraction defined during the provisioning phase.
Then, this update is posted to the PLASE in order to keep track of the PHY topology changes.
At this point, the OONC tries to re-establish the traffic lost due to the hard-failed link.
This is done by repeating the previously described deployment procedure for all the lost LPs, using the same SRC/DST node pairs and original requested rates.

It is remarkable that this approach is carried out on a \emph{best effort} basis: it may allocate the re-established LPs on different wavelengths accordingly to the availability on the new physical paths.
In addition, it may even split up the original requested rates in multiple optical channels delivering lower rates accordingly to the maximum modulation format available in the new physical paths.

\section{Experimental Setup}

An experimental setup has been built in LINKS Foundation's photonics laboratory, aiming to demonstrate the feasibility of such modular, open and disaggregated optical network architecture by means of a proof of concept.
The experimental setup emulating an optical network is depicted in Fig.~\ref{fig:setup} and it is composed by three nodes, each equipped with commercial TRXs and ROADMs and connected by three different multi-span amplified OLSs, obtaining different optical paths for the channels under test~(CUTs).
The TRXs are CFP2-ACO/DCO coherent pluggables from Lumentum, programmed to generate 4 independent signals (DP-QPSK or DP-16QAM modulated) and to continuously monitor the related bit error rate~(BER), providing an updated average value every 15~seconds.
They are plugged in Cassini AS7716-24SC~\cite{oopt} boxes, an open network packet-optical box built by Edgecore that can host line card slots to incorporate ACO/DCO optical ports based on coherent digital signal processing and optical TRXs from leading optical technology partners.
The Cassini whitebox is operated by the OcNOS operating system by IP Infusion, providing configuration and monitoring facilities by means of NETCONF interfaces. 

\begin{figure}[b]
\centering
\includegraphics[width=\linewidth]{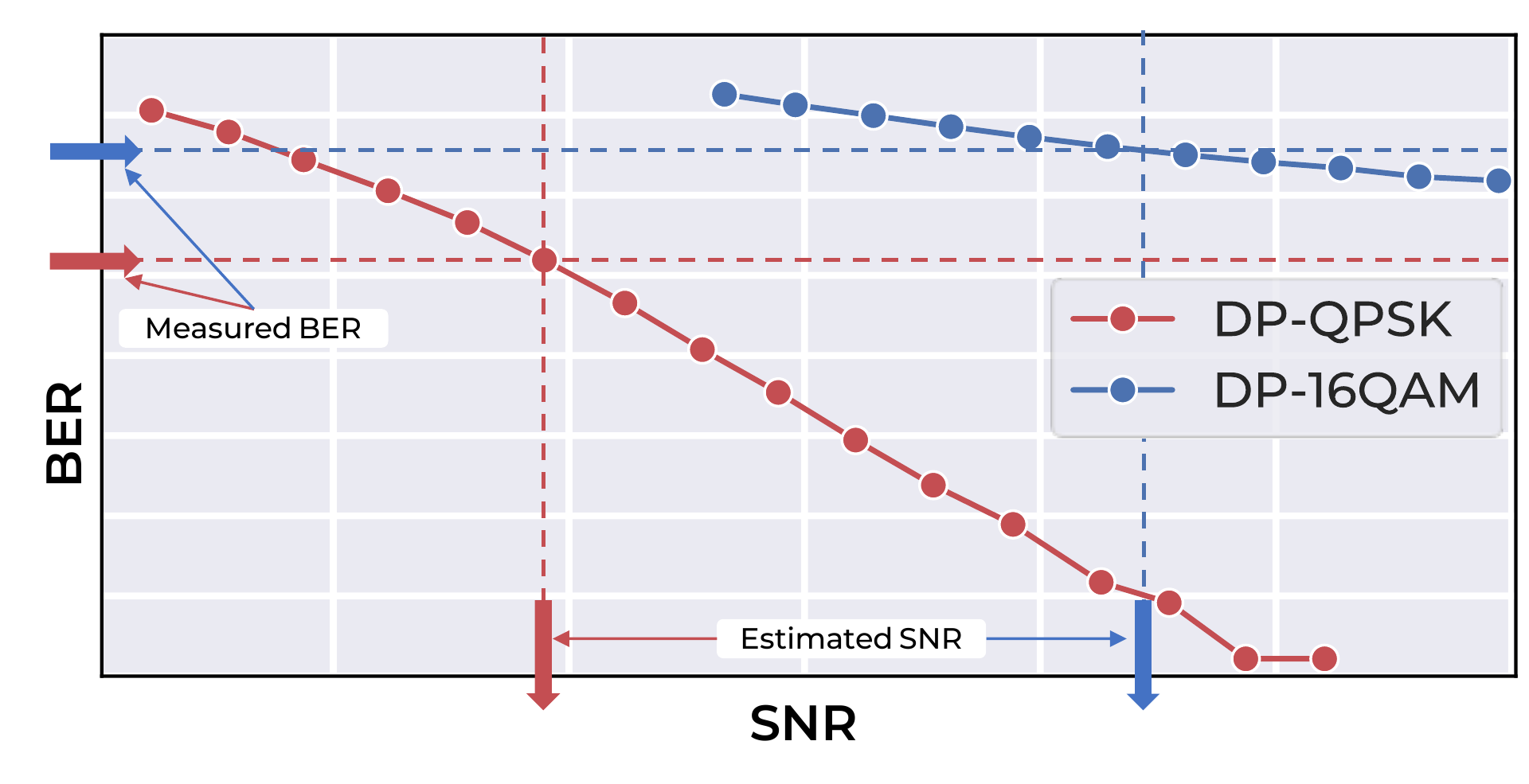}
\caption{SNR estimation procedure using TRX B2B curves given the measured BER.}
\label{fig:ber_vs_snr}
\end{figure}

A C-band wavelength division multiplexing (WDM) comb centered at 193.5~THz and composed by 75~channels, 50~GHz spaced, modulated at 32~GBd each, is generated at the \textit{Node~A} side: 4~CUTs, centered at 192, 193, 194 and 195~THz, respectively, are generated by the TRXs, while a commercial wave shaper filter (1000S from Finisar) is programmed to shape the output of an ASE noise source, generating 71~channels that, coupled with the 4~CUTs, assemble the 75~channels OLS spectral load with no loss of generality because of the large time constant characterizing the physical effects within EDFAs.

\textit{ROADM~1} can be configured to add the 75~channels and to route them towards \textit{Node~C}, either through \textit{Line~1} or \textit{Line~2}: the former straight, LP~1, connects \textit{Node~A} to \textit{Node~C} through 6~spans, each based on commercial EDFA operating in constant gain mode and followed by a standard single mode fiber (SSMF) of 65~km nominal length.
In the middle of the latter, \textit{ROADM~2} can drop the CUTs, so that their BER can be evaluated, or forward them towards \textit{Node~C}.
\textit{Line~2A} and \textit{Line~2B}, composing LP~2, consist of 5~amplified SSMF spans of about 100~km each.
\textit{ROADM~3} finally drops the 4~CUTs either if they are propagated through \textit{Line~1} or \textit{Line~2}.

\begin{table}[t]
\centering
\caption{Physical Layer Characterization Results:\\Complete Set of Retrieved Parameters}
\label{tab:characterization}
\begin{tabular}{|c|c|c|c|c|c|c|}
\hline
\textbf{LINE} &
  \textbf{SPAN} &
  \textbf{\begin{tabular}[c]{@{}c@{}}$\pmb{L_S}$\\ {[}km{]}\end{tabular}} &
  \textbf{\begin{tabular}[c]{@{}c@{}}$\pmb{C_R}$\\ {[}1/W/km{]}\end{tabular}} &
  \textbf{\begin{tabular}[c]{@{}c@{}}$\pmb{D}$\\ {[}ps/nm/km{]}\end{tabular}} &
  \textbf{\begin{tabular}[c]{@{}c@{}}$\pmb{l(0)}$\\ {[}dB{]}\end{tabular}} &
  \textbf{\begin{tabular}[c]{@{}c@{}}$\pmb{l(L_S)}$\\ {[}dB{]}\end{tabular}} \\ \hline
\multirow{6}{*}{1}  & 1  & 65.5  & 0.34 & 16.6 & 5.5 & 0.1 \\ \cline{2-7} 
                    & 2  & 65.3  & 0.34 & 16.8 & 1.4 & 0.3 \\ \cline{2-7} 
                    & 3  & 65.5  & 0.44 & 16.7 & 1.6 & 0.1 \\ \cline{2-7} 
                    & 4  & 65.6  & 0.34 & 16.7 & 0.2 & 1.4 \\ \cline{2-7} 
                    & 5  & 65.2  & 0.42 & 16.7 & 0.5 & 0.4 \\ \cline{2-7} 
                    & 6  & 65.8  & 0.34 & 16.5 & 0.1 & 1.3 \\ \hline
\multirow{5}{*}{2A} & 1  & 106.2 & 0.34 & 17.5 & 3.6 & 0.2 \\ \cline{2-7} 
                    & 2  & 107.5 & 0.44 & 17.9 & 1.2 & 0.7 \\ \cline{2-7} 
                    & 3  & 106.2 & 0.44 & 17.7 & 1.5 & 0.1 \\ \cline{2-7} 
                    & 4  & 108.8 & 0.42 & 17.7 & 0.6 & 0.1 \\ \cline{2-7} 
                    & 5  & 108.3 & 0.42 & 17.8 & 0.2 & 0.1 \\ \hline
\multirow{5}{*}{2B} & 1  & 106.2 & 0.42 & 17.9 & 1.1 & 0.2 \\ \cline{2-7} 
                    & 2  & 106.8 & 0.34 & 17.7 & 0.1 & 0.1 \\ \cline{2-7} 
                    & 3  & 106.4 & 0.34 & 17.7 & 0.2 & 0.7 \\ \cline{2-7} 
                    & 4  & 107.3 & 0.42 & 17.8 & 0.2 & 0.1 \\ \cline{2-7} 
                    & 5 & 108.3 & 0.42 & 17.8 & 0.5 & 2.3 \\ \hline
\end{tabular}
\end{table}

\begin{figure*}[b]
\centering
\includegraphics[width=\linewidth]{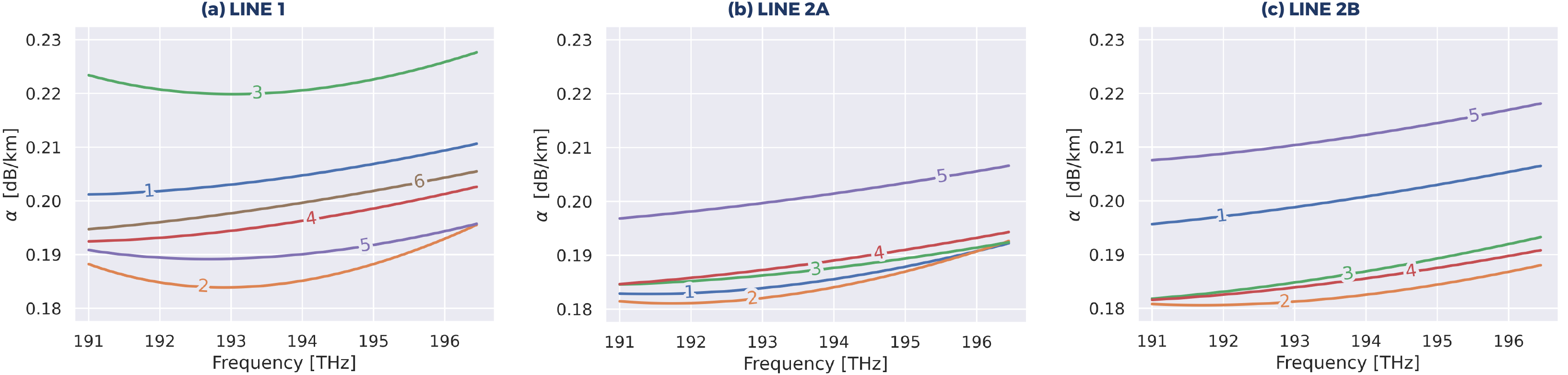}
\caption{Physical layer characterization results: retrieved loss coefficient functions for each optical fiber span within the network set in laboratory.}
\label{fig:characterization}
\end{figure*}

The software implementation includes ONOS, version 2.7.0, as NOS with some additional custom features developed for this proof of concept.
For the first time, ONOS is capable to provide the frequency slot occupation per link with a granularity of 12.5~GHz using new custom REST endpoints.
Moreover, specific drivers have been developed enabling the control of the TRXs through the OcNOS operating system, including the configuration of the desired modulation format.
%
For this proof of concept, a QoT-driven approach is adopted, aiming to favour a vendor-agnostic network management.
As a QoT estimator, GNPy represents the PHY-DT of the control system intelligence, providing the optical propagation model within the PLASE.
ONOS and the PLASE are hosted in two different servers in order to emulate the cloud environment.
ONOS's server is equipped with Intel(R) Pentium(R) CPU G860 @3.00~GHz and 16~GB of RAM.
The PLASE runs in a server powered by Intel(R) Core(TM) i7-4980HQ CPU @2.80~GHz and 16~GB of RAM.
%
The OONC is implemented as a Python framework orchestrating the other software modules, exposing multiple REST endpoints developed based on the Flask library.
%
Each OLC exploits a Secure Shell (SSH) protocol, that allows to open a control flow enabling to set and poll the EDFA's working parameters (e.g. gain, tilt) and performance monitors (e.g. OCM output, total optical input and output power).

\begin{table}[t]
\centering
\caption{EDFA Optimal Working Point}
\label{tab:optimization}
\begin{tabular}{|c|c|c|c|c|}
\hline
\textbf{LINE} &
  \textbf{AMPLIFIER} &
  \textbf{\begin{tabular}[c]{@{}c@{}}G\\ {[}dB{]}\end{tabular}} &
  \textbf{\begin{tabular}[c]{@{}c@{}}T\\ {[}dB{]}\end{tabular}} &
  \textbf{\begin{tabular}[c]{@{}c@{}}P\textsubscript{OUT}\\ {[}dBm{]}\end{tabular}} \\ \hline
\multirow{7}{*}{1}  & BST   & --   & --   & 21.8 \\ \cline{2-5} 
                    & ILA 1 & 15.0 & -0.1 & --   \\ \cline{2-5} 
                    & ILA 2 & 15.0 & -1.4 & --   \\ \cline{2-5} 
                    & ILA 3 & 15.0 & 0.0  & --   \\ \cline{2-5} 
                    & ILA 4 & 15.0 & 0.6  & --   \\ \cline{2-5} 
                    & ILA 5 & 15.7 & -1.0 & --   \\ \cline{2-5} 
                    & PRE   & --   & --   & 20.0 \\ \hline
\multirow{6}{*}{2A} & BST   & --   & --   & 21.8 \\ \cline{2-5} 
                    & ILA 1 & 23.3 & -5.0 & --   \\ \cline{2-5} 
                    & ILA 2 & 22.1 & -5.0 & --   \\ \cline{2-5} 
                    & ILA 3 & 21.6 & -1.9 & --   \\ \cline{2-5} 
                    & ILA 4 & 22.9 & -1.0 & --   \\ \cline{2-5} 
                    & PRE   & --   & --   & 23.0 \\ \hline
\multirow{6}{*}{2B} & BST   & --   & --   & 19.2 \\ \cline{2-5} 
                    & ILA 1 & 22.0 & -5.0 & --   \\ \cline{2-5} 
                    & ILA 2 & 22.2 & -4.8 & --   \\ \cline{2-5} 
                    & ILA 3 & 23.3 & -1.9 & --   \\ \cline{2-5} 
                    & ILA 4 & 23.0 & -1.4 & --   \\ \cline{2-5} 
                    & PRE   & --   & --   & 20.0 \\ \hline
\end{tabular}
\end{table}

\begin{table*}[b]
\centering
\caption{Network Transmission Performance Validation Results}
\label{tab:results}
\begin{tabular}{|cc|cccc|cccc|}
\hline
\multicolumn{2}{|c|}{\multirow{2}{*}{}} &
  \multicolumn{4}{c|}{\textbf{LP 1 (A -\textgreater C)}} &
  \multicolumn{4}{c|}{\textbf{LP 2 (A -\textgreater B -\textgreater C)}} \\ \cline{3-10} 
\multicolumn{2}{|c|}{} &
  \multicolumn{1}{c|}{\textbf{\begin{tabular}[c]{@{}c@{}}CUT 1\\ DCO\\ (192 THz)\end{tabular}}} &
  \multicolumn{1}{c|}{\textbf{\begin{tabular}[c]{@{}c@{}}CUT 2 \\ ACO\\ (193 THz)\end{tabular}}} &
  \multicolumn{1}{c|}{\textbf{\begin{tabular}[c]{@{}c@{}}CUT 3 \\ DCO\\ (194 THz)\end{tabular}}} &
  \textbf{\begin{tabular}[c]{@{}c@{}}CUT 4 \\ ACO\\ (195 THz)\end{tabular}} &
  \multicolumn{1}{c|}{\textbf{\begin{tabular}[c]{@{}c@{}}CUT 1 \\ DCO\\ (192 THz)\end{tabular}}} &
  \multicolumn{1}{c|}{\textbf{\begin{tabular}[c]{@{}c@{}}CUT 2 \\ ACO\\ (193 THz)\end{tabular}}} &
  \multicolumn{1}{c|}{\textbf{\begin{tabular}[c]{@{}c@{}}CUT 3 \\ DCO\\ (194 THz)\end{tabular}}} &
  \textbf{\begin{tabular}[c]{@{}c@{}}CUT 4 \\ ACO\\ (195 THz)\end{tabular}} \\ \hline
\multicolumn{2}{|c|}{\textbf{\begin{tabular}[c]{@{}c@{}}GNPy Prediction\\ {[}dB{]}\end{tabular}}} &
  \multicolumn{1}{c|}{24.0} &
  \multicolumn{1}{c|}{23.7} &
  \multicolumn{1}{c|}{23.7} &
  23.6 &
  \multicolumn{1}{c|}{18.4} &
  \multicolumn{1}{c|}{17.8} &
  \multicolumn{1}{c|}{18.1} &
  17.6 \\ \hline
\multicolumn{1}{|c|}{\multirow{3}{*}{\textbf{\begin{tabular}[c]{@{}c@{}}DP-QPSK\\ (100G)\end{tabular}}}} &
  \textbf{BER} &
  \multicolumn{1}{c|}{1.6e-8} &
  \multicolumn{1}{c|}{9.5e-8} &
  \multicolumn{1}{c|}{1.2e-08} &
  8.6e-08 &
  \multicolumn{1}{c|}{4.2e-05} &
  \multicolumn{1}{c|}{1.9e-04} &
  \multicolumn{1}{c|}{3.5e-05} &
  1.4e-04 \\ \cline{2-10} 
\multicolumn{1}{|c|}{} &
  \textbf{\begin{tabular}[c]{@{}c@{}}GSNR\\ {[}dB{]}\end{tabular}} &
  \multicolumn{1}{c|}{27.1} &
  \multicolumn{1}{c|}{24.6} &
  \multicolumn{1}{c|}{27.5} &
  24.7 &
  \multicolumn{1}{c|}{19.1} &
  \multicolumn{1}{c|}{17.7} &
  \multicolumn{1}{c|}{19.2} &
  18.0 \\ \cline{2-10} 
\multicolumn{1}{|c|}{} &
  \textbf{\begin{tabular}[c]{@{}c@{}}Margin\\ {[}dB{]}\end{tabular}} &
  \multicolumn{1}{c|}{3.1} &
  \multicolumn{1}{c|}{0.9} &
  \multicolumn{1}{c|}{3.7} &
  1.1 &
  \multicolumn{1}{c|}{0.6} &
  \multicolumn{1}{c|}{0.0} &
  \multicolumn{1}{c|}{1.1} &
  0.4 \\ \hline
\multicolumn{1}{|c|}{\multirow{3}{*}{\textbf{\begin{tabular}[c]{@{}c@{}}DP-16QAM\\ (200G)\end{tabular}}}} &
  \textbf{BER} &
  \multicolumn{1}{c|}{3.9e-03} &
  \multicolumn{1}{c|}{9.9e-3} &
  \multicolumn{1}{c|}{4.2e-03} &
  1.1e-02 &
  \multicolumn{1}{c|}{--} &
  \multicolumn{1}{c|}{--} &
  \multicolumn{1}{c|}{--} &
  -- \\ \cline{2-10} 
\multicolumn{1}{|c|}{} &
  \textbf{\begin{tabular}[c]{@{}c@{}}GSNR\\ {[}dB{]}\end{tabular}} &
  \multicolumn{1}{c|}{26.3} &
  \multicolumn{1}{c|}{25.0} &
  \multicolumn{1}{c|}{26.0} &
  24.7 &
  \multicolumn{1}{c|}{--} &
  \multicolumn{1}{c|}{--} &
  \multicolumn{1}{c|}{--} &
  -- \\ \cline{2-10} 
\multicolumn{1}{|c|}{} &
  \textbf{\begin{tabular}[c]{@{}c@{}}Margin\\ {[}dB{]}\end{tabular}} &
  \multicolumn{1}{c|}{2.3} &
  \multicolumn{1}{c|}{1.3} &
  \multicolumn{1}{c|}{2.2} &
  1.0 &
  \multicolumn{1}{c|}{--} &
  \multicolumn{1}{c|}{--} &
  \multicolumn{1}{c|}{--} &
  -- \\ \hline
\end{tabular}
\end{table*}

\section{Results}

In the following, all the experimental results related to the validation of the network optical transmission and the LP-recovery use-case are reported and commented, illustrating the relevant observations and details in terms of practical implementation.
Both the PHY characterization and OLS control optimization methodologies adopted in this work are taken from~\cite{borraccini2021cognitive}.

The experimental setup presents two different models of erbium-doped fiber In-Line Amplifier (ILA) according to the nominal length of the specific fiber span (65 or 100~km) and a BST and a PRE integrated in the Lumentum ROADMs.
Preliminarily, both the ILA models have been characterized at full spectral load (C-band) in constant gain mode varying the gain and tilt parameters with different values of total input power.
Each collected dataset is used to train a couple of artificial neural networks, abstracting and modelling the behaviour of the gain profile and the introduced ASE noise profile of the specific amplifier model.
The ROADM's BST and PRE have been similarly characterized at full spectral load in constant output power mode for different output power values varying the total power of the input C-band spectrum.
The software abstraction of these components is obtained for both the applied gain and introduced ASE noise linearly interpolating in logarithmic units the measured quantities.
%
Both the TRX types (ACO/DCO) have been characterized in the back-to-back (B2B) to obtain the BER vs. SNR curves and consequently retrieving the related SNR threshold assuming $10^{-2}$ as pre-FEC (forward error correction) BER threshold for each available modulation format.
In order to measure the GSNR, as graphically explained in Fig.~\ref{fig:ber_vs_snr}, the used method is to translate the measured BER from the Cassini by means of the B2B characterization obtaining the corresponding SNR~\cite{borraccini2020using}.

\subsection{Physical Layer Characterization}

The measurement process bringing to the definition of the PHY topology starts with an OTDR analysis, performed for each fiber span measuring the fiber span length, $L_S$, and the positions of eventual lumped losses, $l(z)$, present along the specific span.
After that, the BST and each in-line EDFA are set in ASE mode providing at the corresponding output a full C-band ASE spectrum with arbitrary shape.
The latter is measured by OCMs at both terminals of each fiber span.
The two ASE power levels are defined accordingly to the characteristics of the installed apparatus, such as EDFAs' maximum total output power and fiber span total losses, and to the telemetry sensitivity.

The PLASE characterizes each fiber span through an optimization strategy that aims to reproduce the experimental measurements using the PHY optical propagation model.
The PHY parameters to estimate for a single fiber span are the Raman efficiency scale factor, $C_R$, the loss coefficient function, $\alpha(f)$, the input, $l(0)$, and the output connector losses, $l(L_S)$, and the eventual lumped losses detected by the OTDR, $l(0<z<L_S)$.

The results of the PHY characterization are shown in Tab.~\ref{tab:characterization} and Fig.~\ref{fig:characterization}.
Given a specific OLS within the optical network, each fiber span is characterized in terms of the PHY parameters described above.
In addition, the dispersion coefficient, $D$, has been measured for each span before the complete installation of the equipment.
As per the data-sheet, the equivalent representations retrieved by the characterization process correspond to the properties of the SSMF type.

\subsection{OLS Control \& Transmission Performance}

On the basis of the PHY topology, the PLASE optimizes the amplifier working point feeding the PHY model with the retrieved parameters.
The result of the optimization process is reported in Tab.~\ref{tab:optimization}.
The optimization algorithm aims to homogeneously maximize the GSNR along the whole band considering the propagation model of the fibers and amplifiers.
All the amplifiers integrated within the ROADMs (BSTs and PREs) are set to work in constant output power mode.
Instead, the in-line amplifiers are set in constant gain mode.
The adjustment of the amplifier's working point takes place by modifying the device setting parameters, such as total output power, gain or tilt.

The evaluation of the network transmission performance obtained following the described cognitive approach is operated by setting the amplifier working point and estimating the GSNR of each CUT for both the LPs on the base of the measured BER.
Each GSNR estimation is compared to the GSNR value predicted by GNPy, determining the related resulting margin.
The latter is calculated without considering any other contribution, as if the system worked at zero margin.
The summary of the experimental measurement campaign is presented in Tab.~\ref{tab:results}.
The measure of the BER using the DP-16QAM modulation format has been possible only for the case of the short path, LP~1.
Observing the results, it is remarkable the fact that DCO TRXs provide larger margins than ACO ones, presenting a higher intrinsic device robustness.
Comparing the results inherent to the short path, LP~1, for the different modulation formats, it is noted that the estimated GSNRs for the ACOs are comparable, as can be expected given that the degradation introduced by the network does not change.
Instead, there is a non negligible variation for the DCO TRXs changing the modulation format.
Since the QoT is very high for these TRXs, this can be explained by the working point location within the B2B curve, which is affected by a greater uncertainty during the characterization phase.
The GNPy GSNR prediction is conservative in all the cases, obtaining a satisfactory result by working with zero margin.
%
%
Given these considerations, the degree of flatness of the measured QoT also corresponds to the prediction on the basis of the optimization criterion.

\subsection{Lightpath Deployment \& Recovery}

\begin{figure}[t]
\centering
\includegraphics[width=\linewidth]{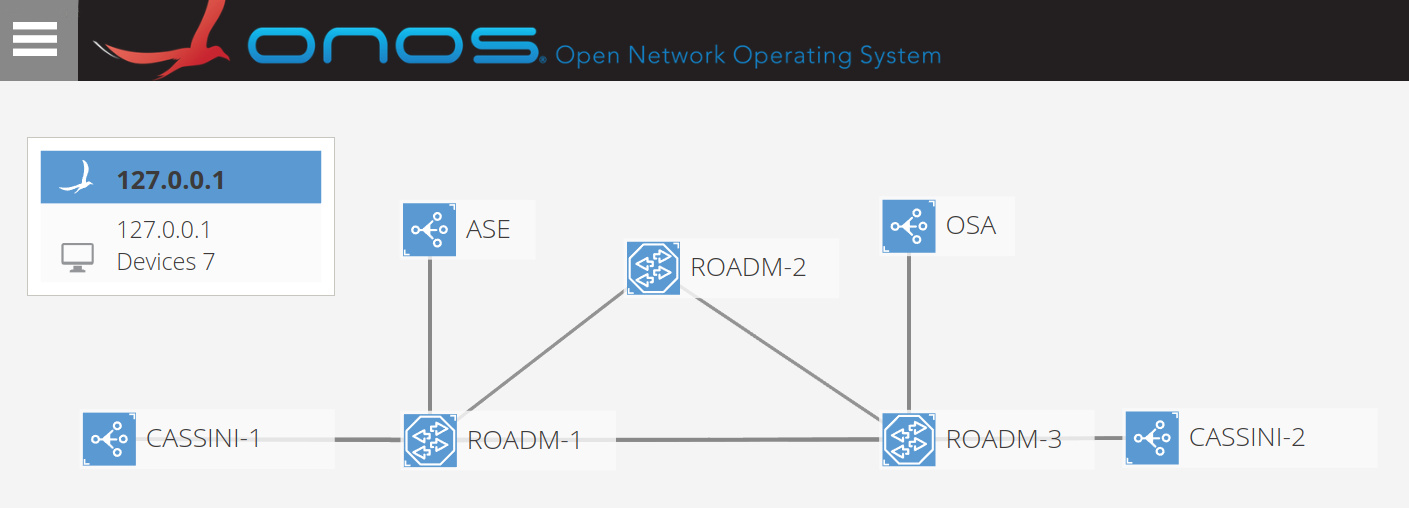}
\caption{ONOS topology (optical nodes and links) of the experimental setup.}
\label{fig:network_topology_onos}
\end{figure}

In order to test the network behaviour, a use-case experiment is conducted following the three main steps described in Fig.~\ref{fig:data_plane}.
During the boot phase, ONOS instantiates a control plane connection with all the physical devices (i.e., 2 Cassini boxes and 3 Lumentum ROADMs).
Fig.~\ref{fig:network_topology_onos} is a screenshot of the the ONOS GUI showing all connected devices and the considered optical links.
Two emulated devices are included (i.e., ASE/OSA), acting as logical termination points of the 71~ASE-shaped channels.
Given the topology and spectral occupancy information provided by ONOS, thanks to the custom developed ONOS endpoints, the OONC is able to summarize the network status in terms of physical paths and free channels, initializing the routing space.

\begin{figure}[t]
\centering
\includegraphics[width=\linewidth]{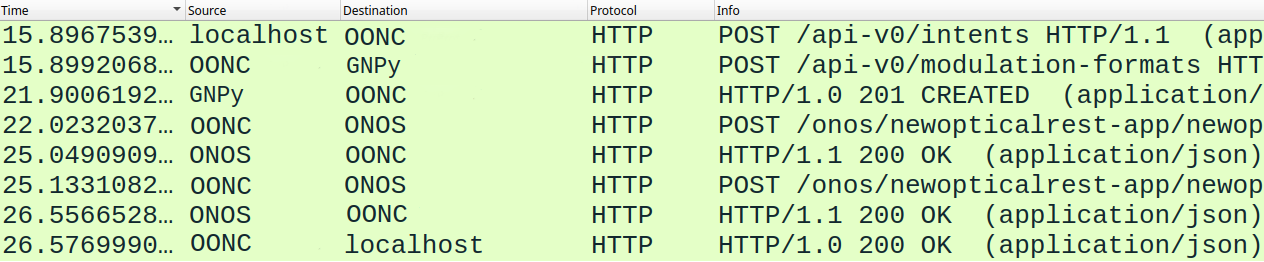}
\caption{REST requests exchanged between the software modules to satisfy a generic traffic request coming from the application/IP layer (by Wireshark).}
\label{fig:wireshark_deploy}
\end{figure}

For the considered use-case, a traffic request of 400G between the nodes A and C is generated and sent to the OONC.
The PLASE replies with the modulation formats computed for each channel frequency for all the paths having the same SRC and DST nodes. 
The OONC carries out the RSA algorithm to select the available channels with the highest available modulation format cardinality.
The chosen LPs are sent to ONOS through the developed REST interfaces, which operatively deploy the LP on the devices.
Based on the transmission performance simulation shown in Tab.~\ref{tab:results}, the traffic request is satisfied with two DP-16QAM on the shorter line, Line~1.
All the exchanged REST requests regarding the deployment of the LPs satisfying the given traffic request are described in Fig.~\ref{fig:wireshark_deploy}.

\begin{figure}[t]
\centering
\includegraphics[width=\linewidth]{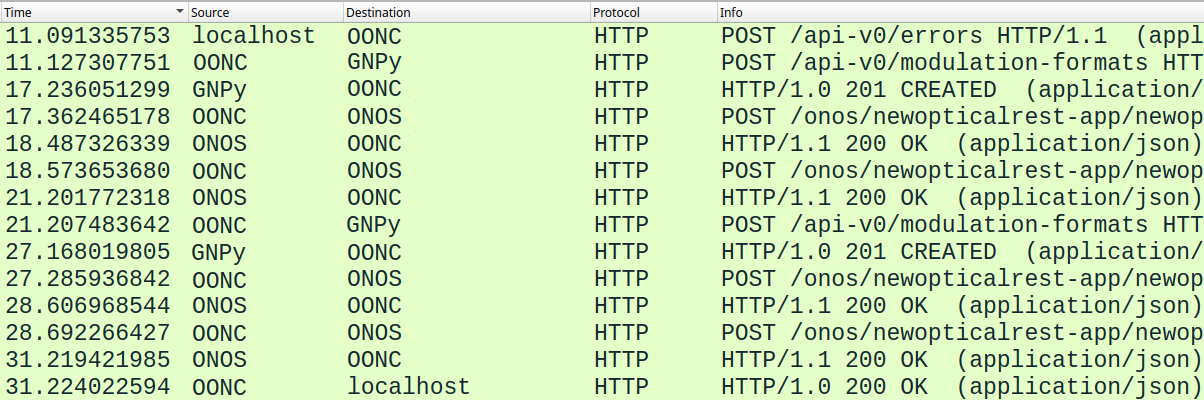}
\caption{REST requests exchanged between the software modules to recover the lost traffic after a hard-failure (by Wireshark).}
\label{fig:wireshark_recovery}
\end{figure}

As last step, a hard failure is emulated on Line~1 with the two active connections.
Once ONOS notifies to the OONC the change of the PHY, the topology abstraction is updated and the lost traffic is estimated. 
Thus, the LP deployment procedure is triggered on the updated topology, trying to recover the lost traffic between the nodes A and C.
As the DP-16QAM is not feasible along the longer line, in order to recover the whole 400G lost traffic, 4x DP-QPSK intents are chosen.
All the messages exchanged during the LP recovery process are shown in Fig.~\ref{fig:wireshark_recovery}.

\begin{table}[b]
\centering
\caption{Lightpath Recovery Time Lapse Measurements}
\label{tab:recovery_results}
\begin{tabular}{|c|c|}
\hline
\textbf{Interaction} & \textbf{Time [sec]} \\ \hline
Topology Update & 0.017 \\ \hline
Lost Traffic Estimation & 0.012 \\ \hline
L-PCE & 6.580 \\ \hline
Lightpath Establishment & 4.870 \\ \hline
\textbf{Total Recovery} & \textbf{11.708} \\ \hline
\end{tabular}
\end{table}

The time lapse measurements regarding the control interactions performing the LP recovery are summarized in Tab.~\ref{tab:recovery_results}. 
These measurements refer to a single 200G LP recovery (2x DP-QPSK deployed LPs), which are comparable with the initial case of 400G LP deployment (2x DP-16QAM deployed LPs).
The two most significant time-consuming processes are the L-PCE and LP establishment, spending roughly 4.5 and 6.5~seconds, respectively.
The first one is a CPU-bound process where the dominant contribution is given by the computation of the various artificial neural networks simulating the optical amplifier behaviour.
Because of this, a reduction of roughly one order of magnitude in the execution time is expected using a dedicated hardware and a further software optimization.
The LP establishment process is consistent with other experiments available in literature, where most of the time is required by the specific device to set the connection~\cite{giorgetti2020control}.
Also this value can be decreased by means of hardware updates and next generation transceivers.

\section{Conclusion}

An architecture for the control of an open and disaggregated optical network has been presented in terms of the players involved and their interactions, highlighting the independent operation of the optical control and data planes.
As a proof of concept, a multi-vendor triangular-topology optical network experimental setup has been built to verify the suggested network architecture.
The main adopted management strategies are the use of the optical network at full spectral load and a QoT-driven cognitive vendor-agnostic process of the control plane with regards to the characterization of the physical layer and the optimization of the amplifiers' working point.
The implementation of the software framework involves ONOS as network operative system and GNPy as physical layer digital twin.
The tested features of the proof of concept were the transmission performance of the two LPs in the topology as well as the operation of the deployment and recovery procedure through relative time lapse measurements, obtaining deployment time intervals compatible with recent analogous works.

\section*{Acknowledgments}
The authors wish to thank Lumentum, Telecom Infra Project, Cisco, IP Infusion, Open Network Foundation for providing hardware and software.
%


\bibliographystyle{IEEEtran}
\bibliography{references.bib}

\begin{thebibliography}{10}
\providecommand{\url}[1]{#1}
\csname url@samestyle\endcsname
\providecommand{\newblock}{\relax}
\providecommand{\bibinfo}[2]{#2}
\providecommand{\BIBentrySTDinterwordspacing}{\spaceskip=0pt\relax}
\providecommand{\BIBentryALTinterwordstretchfactor}{4}
\providecommand{\BIBentryALTinterwordspacing}{\spaceskip=\fontdimen2\font plus
\BIBentryALTinterwordstretchfactor\fontdimen3\font minus
  \fontdimen4\font\relax}
\providecommand{\BIBforeignlanguage}[2]{{%
\expandafter\ifx\csname l@#1\endcsname\relax
\typeout{** WARNING: IEEEtran.bst: No hyphenation pattern has been}%
\typeout{** loaded for the language `#1'. Using the pattern for}%
\typeout{** the default language instead.}%
\else
\language=\csname l@#1\endcsname
\fi
#2}}
\providecommand{\BIBdecl}{\relax}
\BIBdecl

\bibitem{facebook}
``https://code.facebook.com/posts/1977308282496021/an-open-approach-for-switching-routing-and-transport/.''

\bibitem{riccardi-jlt18}
E.~Riccardi, P.~Gunning, O.~G. de~Dios, M.~Quagliotti, V.~López, and A.~Lord,
  ``An operator view on the introduction of white boxes into optical
  networks,'' \emph{Journal of Lightwave Technology}, vol.~36, no.~15, pp.
  3062--3072, 2018.

\bibitem{yang}
M.~Bjorklund, ``{YANG} - a data modeling language for the network configuration
  protocol ({NETCONF}),'' IETF RFC 6020.

\bibitem{dallaglio-jocn17}
M.~Dallaglio, N.~Sambo, F.~Cugini, and P.~Castoldi, ``Control and management of
  transponders with {NETCONF} and {YANG},'' \emph{IEEE/OSA JOCN}, vol.~9,
  no.~3, pp. B43--B52, March 2017.

\bibitem{rfc-netconf}
R.~Enns, M.~Bjorklund, J.~Schoenwaelder, and A.~Bierman, ``Network
  configuration protocol {(NETCONF)},'' IETF RFC 6241, June 2011.

\bibitem{openconfig}
``http://www.openconfig.net.''

\bibitem{open-roadm}
``http://www.openroadm.org.''

\bibitem{tip}
``https://telecominfraproject.com/.''

\bibitem{t600}
``https://www.fujitsu.com/us/products/network/products/1finity-t600/.''

\bibitem{sambo-jlt19}
N.~Sambo, K.~Christodoulopoulos, N.~Argyris, P.~Giardina, C.~Delezoide,
  A.~Sgambelluri, A.~Kretsis, G.~Kanakis, F.~Fresi, G.~Bernini,
  H.~Avramopoulos, E.~Varvarigos, and P.~Castoldi, ``Experimental demonstration
  of a fully disaggregated and automated white box comprised of different types
  of transponders and monitors,'' \emph{Journal of Lightwave Technology},
  vol.~37, no.~3, pp. 824--830, 2019.

\bibitem{sambo-trial19}
N.~Sambo, K.~Christodoulopoulos, N.~Argyris, P.~Giardina, C.~Delezoide,
  D.~Roccato, A.~Percelsi, R.~Morro, A.~Sgambelluri, A.~Kretsis, G.~Kanakis,
  G.~Bernini, E.~Varvarigos, and P.~Castoldi, ``Field trial: Demonstrating
  automatic reconfiguration of optical networks based on finite state
  machine,'' \emph{Journal of Lightwave Technology}, vol.~37, no.~16, pp.
  4090--4097, 2019.

\bibitem{sgambelluri-access20}
A.~Sgambelluri, A.~Giorgetti, D.~Scano, F.~Cugini, and F.~Paolucci,
  ``Openconfig and openroadm automation of operational modes in disaggregated
  optical networks,'' \emph{IEEE Access}, vol.~8, pp. 190\,094--190\,107, 2020.

\bibitem{velasco-jlt18}
L.~Velasco, A.~Sgambelluri, R.~Casellas, L.~Gifre, J.-L. Izquierdo-Zaragoza,
  F.~Fresi, F.~Paolucci, R.~Martínez, and E.~Riccardi, ``Building autonomic
  optical whitebox-based networks,'' \emph{Journal of Lightwave Technology},
  vol.~36, no.~15, pp. 3097--3104, 2018.

\bibitem{kundrat-jlt18}
J.~Kundrat, J.~Vojtech, P.~Skoda, R.~Vohnout, J.~Radil, and O.~Havlis,
  ``{YANG/NETCONF ROADM}: Evolving open {DWDM Toward SDN} applications,''
  \emph{Journal of Lightwave Technology}, vol.~36, no.~15, pp. 3105--3114,
  2018.

\bibitem{nadal-jocn21}
L.~Nadal, J.~M. Fabrega, M.~S. Moreolo, F.~J. Vilchez, R.~Casellas, R.~Munoz,
  R.~Vilalta, and R.~Martinez, ``Programmable disaggregated multi-dimensional
  s-bvt as an enabler for high capacity optical metro networks,'' \emph{Journal
  of Optical Communications and Networking}, vol.~13, no.~6, pp. C31--C40,
  2021.

\bibitem{paolucci-jlt18}
F.~Paolucci, A.~Sgambelluri, F.~Cugini, and P.~Castoldi, ``Network telemetry
  streaming services in sdn-based disaggregated optical networks,''
  \emph{Journal of Lightwave Technology}, vol.~36, no.~15, pp. 3142--3149,
  2018.

\bibitem{campanella-ofc19}
A.~Campanella, H.~Okui, A.~Mayoral, D.~Kashiwa, O.~G. de~Dios, D.~Verchere,
  Q.~Pham~Van, A.~Giorgetti, R.~Casellas, R.~Morro, and L.~Ong, ``Odtn: Open
  disaggregated transport network. discovery and control of a disaggregated
  optical network through open source software and open apis,'' in \emph{2019
  Optical Fiber Communications Conference and Exhibition (OFC)}, 2019, pp.
  1--3.

\bibitem{filer2018multi}
M.~Filer, M.~Cantono, A.~Ferrari, G.~Grammel, G.~Galimberti, and V.~Curri,
  ``Multi-vendor experimental validation of an open source qot estimator for
  optical networks,'' \emph{Journal of Lightwave Technology}, vol.~36, no.~15,
  pp. 3073--3082, 2018.

\bibitem{oopt-gnpy}
``https://github.com/telecominfraproject/oopt-gnpy.''

\bibitem{curri-ofc21}
E.~Virgillito, R.-P. Braun, D.~Breuer, A.~Gladisch, V.~Curri, and G.~Grammel,
  ``Testing {TIP} open source solutions in deployed optical networks,'' in
  \emph{2021 Optical Fiber Communications Conference and Exhibition (OFC)},
  2021, pp. 1--3.

\bibitem{curri2022gnpy}
V.~Curri, ``Gnpy model of the physical layer for open and disaggregated optical
  networking,'' \emph{Journal of Optical Communications and Networking},
  vol.~14, no.~6, pp. C92--C104, 2022.

\bibitem{borraccini2022qot}
G.~Borraccini, S.~Straullu, A.~Giorgetti, R.~D’Ingillo, D.~Scano,
  A.~D’Amico, E.~Virgillito, A.~Nespola, N.~Sambo, F.~Cugini \emph{et~al.},
  ``Qot-driven optical control and data plane in multi-vendor disaggregated
  networks,'' in \emph{Optical Fiber Communication Conference}.\hskip 1em plus
  0.5em minus 0.4em\relax Optica Publishing Group, 2022, pp. M4F--5.

\bibitem{borraccini2023disaggregated}
G.~Borraccini, R.~Ambrosone, A.~Giorgetti, S.~Straullu, F.~Aquilino,
  E.~Virgillito, A.~D’Amico, R.~D’Ingillo, N.~Sambo, F.~Cugini, and
  V.~Curri, ``Disaggregated optical network orchestration based on the physical
  layer digital twin,'' in \emph{Optical Fiber Communication Conference
  (submitted)}, 2023.

\bibitem{onos}
``https://opennetworking.org/onos/.''

\bibitem{giorgetti2020control}
A.~Giorgetti, A.~Sgambelluri, R.~Casellas, R.~Morro, A.~Campanella, and
  P.~Castoldi, ``Control of open and disaggregated transport networks using the
  open network operating system (onos),'' \emph{Journal of Optical
  Communications and Networking}, vol.~12, no.~2, pp. A171--A181, 2020.

\bibitem{borraccini2021cognitive}
G.~Borraccini, A.~D’Amico, S.~Straullu, A.~Nespola, S.~Piciaccia, A.~Tanzi,
  G.~Galimberti, S.~Bottacchi, S.~Swail, and V.~Curri, ``Cognitive and
  autonomous qot-driven optical line controller,'' \emph{Journal of Optical
  Communications and Networking}, vol.~13, no.~10, pp. E23--E31, 2021.

\bibitem{pham2020demonstration}
Q.~Pham-Van, V.~L{\'o}pez, A.~M. Lopez-de Lerma, R.~Szwedowski, K.~Mr{\'o}wka,
  S.~Auer, H.-T. Thieu, Q.-H. Tran, D.~Verchere, G.~Atkinson \emph{et~al.},
  ``Demonstration of alarm correlation in partially disaggregated optical
  networks,'' in \emph{Optical Fiber Communication Conference}.\hskip 1em plus
  0.5em minus 0.4em\relax Optica Publishing Group, 2020, pp. M3Z--6.

\bibitem{9382009}
A.~D’Amico, S.~Straullu, G.~Borraccini, E.~London, S.~Bottacchi,
  S.~Piciaccia, A.~Tanzi, A.~Nespola, G.~Galimberti, S.~Swail, and V.~Curri,
  ``Enhancing lightpath qot computation with machine learning in partially
  disaggregated optical networks,'' \emph{IEEE Open Journal of the
  Communications Society}, vol.~2, pp. 564--574, 2021.

\bibitem{pointurier2017design}
Y.~Pointurier, ``Design of low-margin optical networks,'' \emph{Journal of
  Optical Communications and Networking}, vol.~9, no.~1, pp. A9--A17, 2017.

\bibitem{zervas2010cognitive}
G.~S. Zervas and D.~Simeonidou, ``Cognitive optical networks: Need,
  requirements and architecture,'' in \emph{2010 12th International Conference
  on Transparent Optical Networks}.\hskip 1em plus 0.5em minus 0.4em\relax
  IEEE, 2010, pp. 1--4.

\bibitem{wei2012cognitive}
W.~Wei, C.~Wang, and J.~Yu, ``Cognitive optical networks: key drivers, enabling
  techniques, and adaptive bandwidth services,'' \emph{IEEE Communications
  magazine}, vol.~50, no.~1, pp. 106--113, 2012.

\bibitem{oopt}
\BIBentryALTinterwordspacing
TIP, ``Open optical \& packet transport project,'' Tech. Rep. [Online].
  Available: \url{https://telecominfraproject.com/oopt/}
\BIBentrySTDinterwordspacing

\bibitem{borraccini2020using}
G.~Borraccini, S.~Straullu, A.~Ferrari, E.~Virgillito, S.~Bottacchi, S.~Swail,
  S.~Piciaccia, G.~Galimberti, G.~Grammel, and V.~Curri, ``Using qot-e for open
  line controlling and modulation format deployment: an experimental proof of
  concept,'' in \emph{2020 European Conference on Optical Communications
  (ECOC)}.\hskip 1em plus 0.5em minus 0.4em\relax IEEE, 2020, pp. 1--4.

\end{thebibliography}

\end{document}